\documentclass[numberedappendix]{emulateapj}

\usepackage[usenames]{color}

\usepackage{rotating}

\shorttitle{The Galactic Center Cloud G0.253+0.016 at High Angular
  Resolution}
\shortauthors{Kauffmann, Pillai \& Zhang}

\begin{document}

\title{The Galactic Center Cloud G0.253+0.016:\\ A Massive Dense Cloud with
  low Star Formation Potential}

\author{Jens Kauffmann, Thushara Pillai\altaffilmark{1}}
\affil{California Institute of Technology, Astronomy Department, 1200
  East California Blvd., Pasadena, CA 91125, USA}
\altaffiltext{1}{CARMA Fellow}

\author{Qizhou Zhang}
\affil{Harvard--Smithsonian Center for Astrophysics, 60 Garden Street
  MS78, Cambridge, MA 02138, USA}

\slugcomment{Copyright 2012 California Institute of Technology. All
  rights reserved. Government sponsorship acknowledged.}

\email{jens.kauffmann@astro.caltech.edu}

\begin{abstract}
  We present the first interferometric molecular line and dust
  emission maps for the Galactic Center (GC) cloud G0.253+0.016,
  observed using the Combined Array for Research in Millimeter--wave
  Astronomy (CARMA) and the Submillimeter Array (SMA). This cloud is
  very dense, and concentrates a mass exceeding the Orion Molecular
  Cloud Complex ($2\times{}10^5\,M_{\sun}$) into a radius of only
  $3~\rm{}pc$, but it is essentially starless. G0.253+0.016 therefore
  violates ``star formation laws'' presently used to explain trends in
  galactic and extragalactic star formation by a factor
  $\sim{}45$. Our observations show a lack of dense cores of
  significant mass and density, thus explaining the low star formation
  activity. Instead, cores with low densities and line widths
  $\lesssim{}1~\rm{}km\,s^{-1}$---probably the narrowest lines
  reported for the GC region to date---are found. Evolution over
  several $10^5~\rm{}yr$ is needed before more massive cores, and
  possibly an Arches--like stellar cluster, could form. Given the
  disruptive dynamics of the GC region, and the potentially unbound
  nature of G0.253+0.016, it is not clear that this evolution will
  happen.
\end{abstract}


\maketitle

\section{Introduction}
\label{sec:introduction}
It is generally understood that the level of ongoing star formation
(SF) activity in a cloud correlates with the reservoir of dense
gas. This concept first became important for extragalactic research
(e.g., \citealt{gao2004:hcn}), and has since been expanded to include
the Milky Way
\citep{wu2010:hcn,lada2010:sf-efficiency,heiderman2010:sf-law,gutermuth2011:sf-law}. Such
work suggests that (\textit{i}) the mass of dense gas and star
formation rate are proportional, and that (\textit{ii}) the
proportionality constant is the same for all clouds near and
far. These are the key results for efforts to understand Milky Way and
extragalactic SF \citep{kennicutt2012:review}.

It is thus interesting to study regions like the Galactic Center (GC)
molecular cloud G0.253+0.016 (or M0.25+0.01:
\citealt{guesten1981:cmz-nh3})---that is more massive and dense than
the Orion~A cloud ($\sim{}2\times{}10^5\,M_{\sun}$ in $2.8~\rm{}pc$
radius for G0.253+0.016; \citealt{lis1994:m0.25};
\citealt{longmore2011:m025}, hereafter L12), but does hardly form
stars at all \citep{lis1994:m0.25}. An infrared luminosity of the
entire cloud of $\le{}3\times{}10^5\,L_{\sun}$, and the absence of
embedded compact H\textsc{ii} regions in $8.4~\rm{}GHz$ VLA maps,
imply $\lesssim{}5$ embedded stars earlier than B0
\citep{lis2001:ir-spectra}. Spitzer can provide more stringent limits,
as it can detect SF at luminosities of a few $10^3\,L_{\sun}$ out to
distances $\approx{}7~\rm{}kpc$ (e.g., in Infrared Dark Clouds
[IRDCs]; \citealt{zhang2009:early-phase-fragmentation} and
\citealt{pillai2011:initial-conditions}). However, this analysis is
beyond the scope of the present paper. A faint $\rm{}H_2O$ maser has
been detected in the cloud \citep{lis1994:m0.25}, but no other masers
reside in the area \citep{caswell2010:gal-center-methanol-masers,
  caswell2011:atca-h20}. G0.253+0.016 appears to be in a very extreme
physical state, with gas kinetic temperatures $\sim{}80~\rm{}K$
exceeding dust temperatures $\le{}30~\rm{}K$
(\citealt{guesten1981:cmz-nh3, carey1998:irdc-properties,
  lis2001:ir-spectra}; L12). G0.253+0.016 forms part of the
$\sim{}100~\rm{}pc$ circumnuclear ring of clouds
\citep{molinari2011:cmz-ring} at $\sim{}8.5~\rm{}kpc$ distance (also
see L12).

Dense GC clouds may play a key role in the mysterious
formation of compact and massive stellar aggregates like the
``Arches'' cluster (\citealt{lis1998:m025}; L12). For all these
reasons, we present the first high--resolution interferometric line
and dust emission data on G0.253+0.016, obtained using CARMA and the
SMA.

\begin{figure*}
\includegraphics[width=\linewidth]{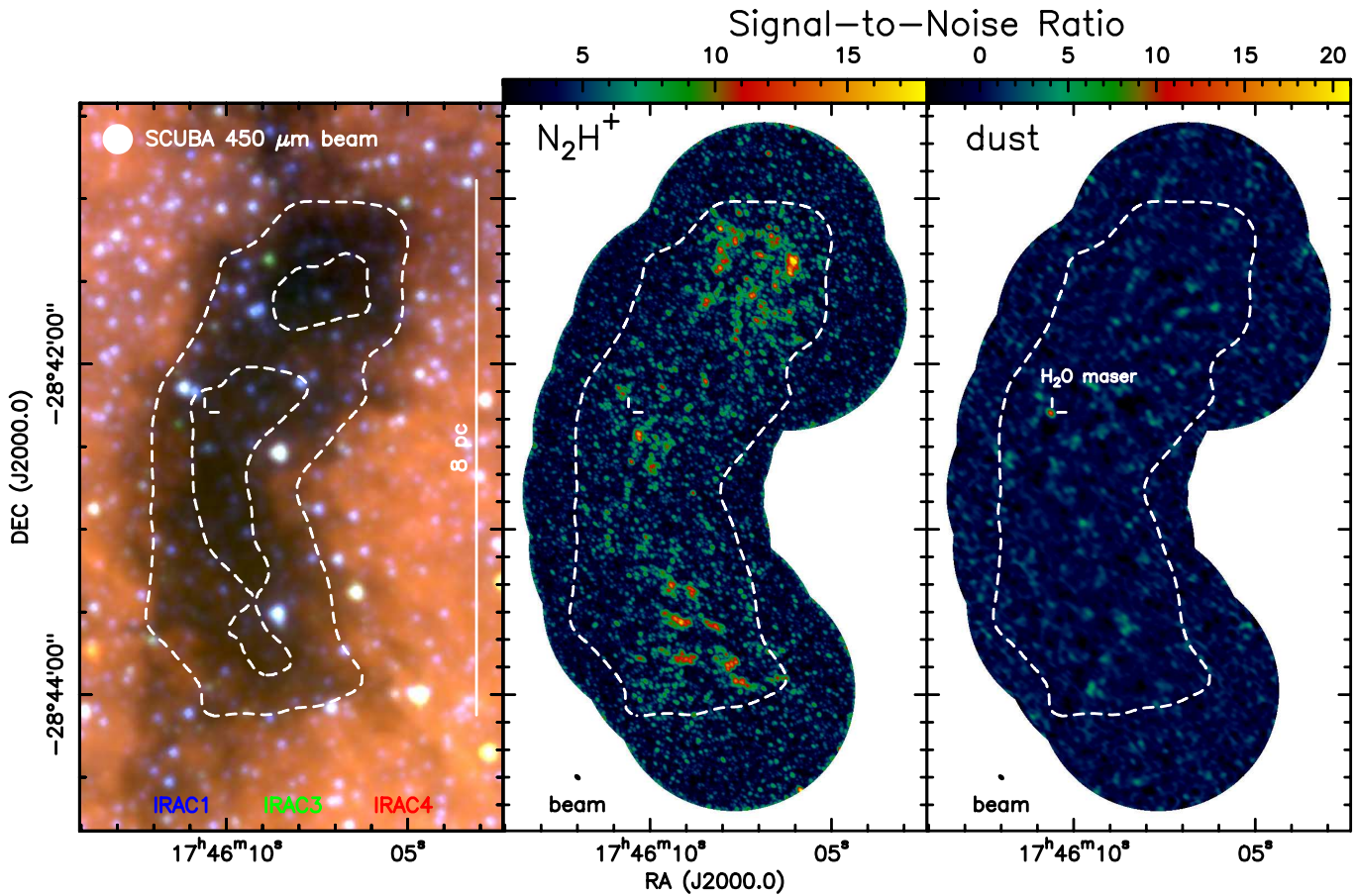}
\caption{Spitzer and SMA maps of G0.253+0.016. The \emph{left} panel
  presents Spitzer IRAC data. Overlaid are $450~\rm{}\mu{}m$
  wavelength intensity contours at 30 and $70~\rm{}Jy\,beam^{-1}$
  (SCUBA Legacy Archive;
  \citealt{difrancesco2008:scuba_catalogue}). The lower contour is
  repeated in all maps shown. The \emph{middle and righ panel} present
  signal--to--noise maps of the $\rm{}N_2H^+$ (3--2) and
  $280~\rm{}GHz$ dust continuum probed by the SMA. The $\rm{}H_2O$
  maser reported by \citet{lis1994:m0.25} is marked.\label{fig:sma}}
\end{figure*}

\section{Observations \& Data Reduction}
\label{sec:observations}

The Submillimeter Array (SMA\footnote {The Submillimeter Array is a
  joint project between the Smithsonian Astrophysical Observatory and
  the Academia Sinica Institute of Astronomy and Astrophysics, and is
  funded by the Smithsonian Institution and the Academia Sinica.})
$\rm{}N_2H^+$ (3--2; $\approx{}0.34~\rm{}km\,s^{-1}$ resolution) and
continuum observations near 280~GHz ($4~\rm{}GHz$ total bandwidth) were made
with seven antennas in compact--north configuration in a single track
in June 2009. Eleven positions separated at less than half a
$42\arcsec$ primary beam were observed. The 345~GHz receiver was tuned
to the $\rm{}N_2H^+$ line in the LSB spectral band s4, using 256
channels per chunk and 24 chunks per sideband. The data were taken
under good weather conditions at $<1.3~\rm{}mm$ water vapor with
characteristic system temperatures $<180~\rm{}K$.

The Combined Array for Research in Millimeter--wave Astronomy
(CARMA\footnote {Support for CARMA construction was derived from the
  Gordon and Betty Moore Foundation, the Kenneth T.\ and Eileen L.\
  Norris Foundation, the James S.\ McDonnell Foundation, the
  Associates of the California Institute of Technology, the University
  of Chicago, the states of California, Illinois, and Maryland, and
  the National Science Foundation (NSF). Ongoing CARMA development and
  operations are supported by the NSF and the CARMA partner
  universities.})  observations were executed in CARMA23 mode in
November 2011 in a combined D and SH configuration. Four USB bands
were used to observe spectral lines ($\rm{}N_2H^+$ [1--0],
$\rm{}HCO^+$ [1--0], SiO [2--1]; $\approx{}0.5~\rm{}km\,s^{-1}$
resolution; $\rm{}HCO^+$ is not analyzed here), and continuum (500~MHz
bandwidth) for calibration purposes. Six positions, spaced at half the
$\approx{}80\arcsec$ primary beam for the 10.4m--telescopes, were
observed. We flagged 3.5m--telescope baselines $<50~\rm{}ns$ to reduce
sidelobes.

Calibration and imaging were done using MIR (an IDL--based SMA package),
MIRIAD, and GILDAS. Flux calibrations using Titan and Uranus for the
SMA, and Neptune for CARMA are expected to be accurate within 20\%.

\begin{figure*}
\centerline{\includegraphics[width=0.8\linewidth]{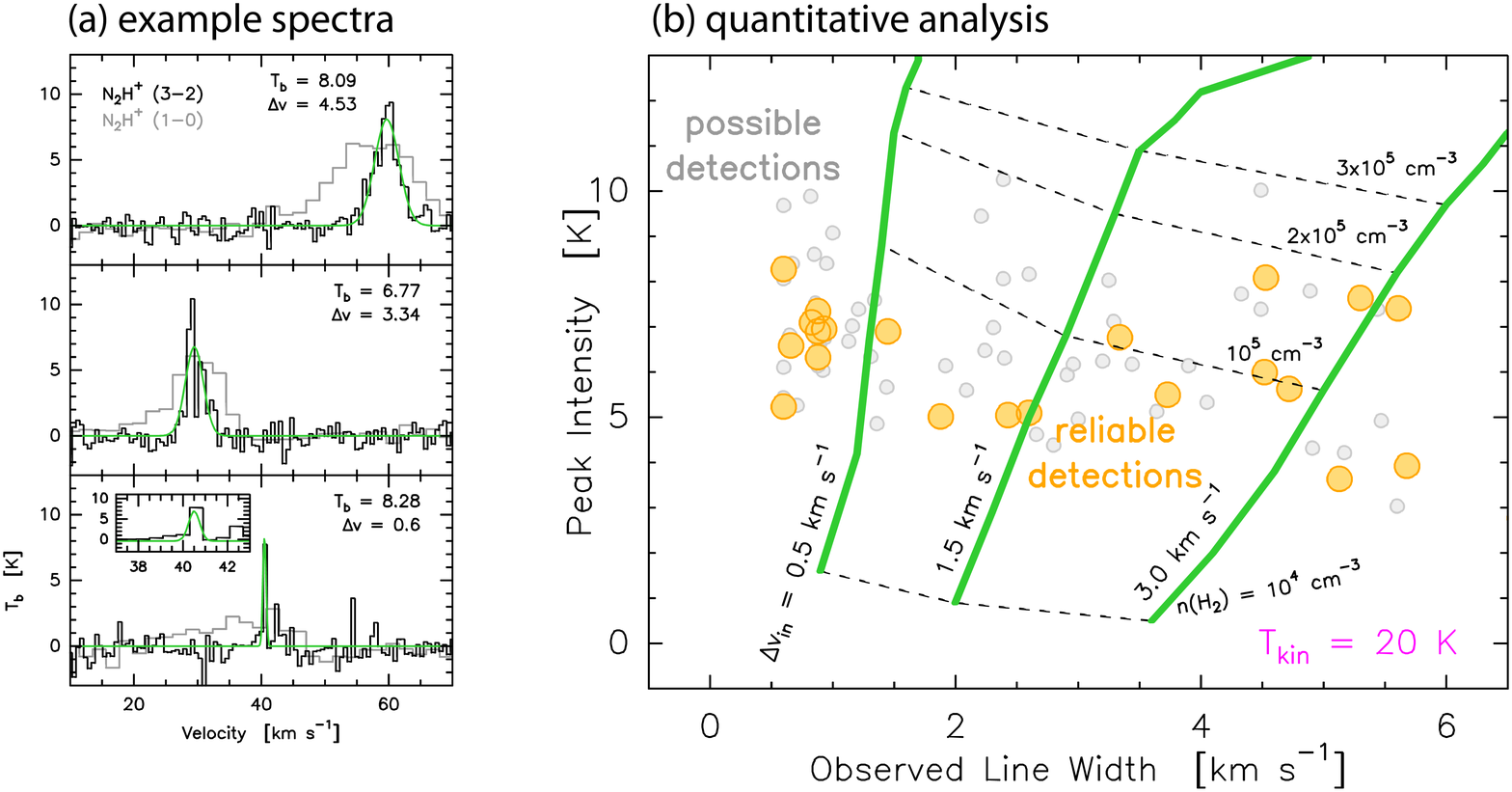}}
\caption{Analysis of SMA $\rm{}N_2H^+$ (3--2) data. \emph{Panel (a)}
  shows three bright reliable example spectra sampling the full range
  of observed velocity dispersions. CARMA $\rm{}N_2H^+$ (1--0) spectra
  are overlaid for reference. Gaussian fits are summarized by \emph{green
  lines and fitted parameters}. These fit results are analyzed in
  \emph{panel (b)}: possible and reliable detections are marked by \emph{grey
  and yellow circles}, respectively. Model line widths and intensities,
  calculated with MOLLIE \citep{keto2010:mollie} using a kinetic
  temperature of $20~\rm{}K$, are indicated (using \emph{green and
    dashed black lines}, respectively) for a range of intrinsic line
  widths, $\Delta{}v_{\rm{}in}$, and densities at $0.1~\rm{}pc$
  radius, $n({\rm{}H_2})$.\label{fig:n2h-analysis}}
\end{figure*}

\section{Results}
\label{sec:results}
\subsection{SMA Dust Emission: No Compact
  Cores\label{sec:dust-densities}}
Figure~\ref{fig:sma}~(right) presents the $280~\rm{}GHz$ dust emission
data, observed with a beam size of $2\farcs{}6\times{}1\farcs{}8$ (PA
$48\degr{}$). A continuum peak of $90~\rm{}mJy\,beam^{-1}$ is detected
within $0\farcs{}5$ of the aforementioned $\rm{}H_2O$ maser position reported
by \citet{lis1994:m0.25}. The remaining part of the map is free of emission above
the $5\sigma$--noise--level of $30~\rm{}mJy\,beam^{-1}$.

We adopt a dust temperature of $20~\rm{}K$, following Herschel--based
estimates of $20$--$25~\rm{}K$ (L12), and
\citet{ossenkopf1994:opacities} dust opacities scaled down by a factor
1.5 ($0.008~\rm{}cm^2\,g^{-1}$; see
\citealt{kauffmann2010:mass-size-i} and Appendix~A of
\citealt{kauffmann2008:mambo-spitzer}). The $5\sigma$--noise--level
corresponds to an $\rm{}H_2$ column density of
$1.7\times{}10^{23}~\rm{}cm^{-2}$. Towards the $\rm{}H_2O$ maser, the
column density derived from the intensity is
$5.2\times{}10^{23}~\rm{}cm^{-2}$. This yields masses per beam of
$<26\,M_{\sun}$ and $78\,M_{\sun}$, respectively, when integrating the
column densities over the the half power beam width (of
$0.046~\rm{}pc$ effective radius).

Note that L12 use \citet{ossenkopf1994:opacities} opacities. For
consistency, we increase their Herschel--based mass measurement
by a factor 1.5.

\begin{figure*}
\begin{center}
\includegraphics[width=\linewidth]{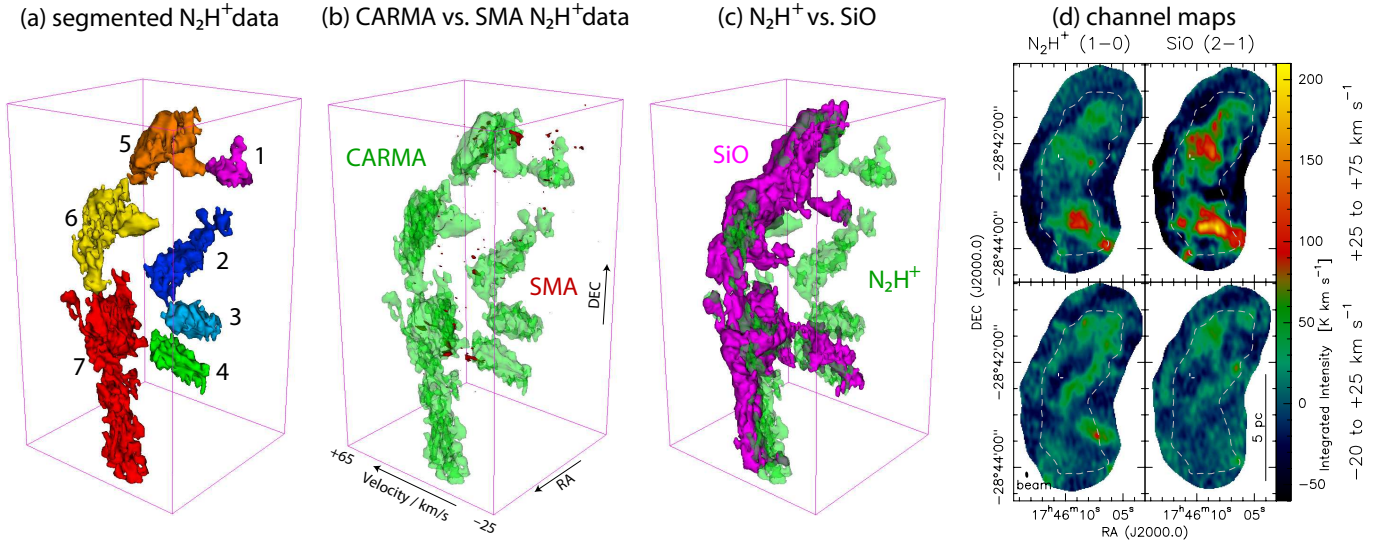}
\end{center}
\caption{CARMA maps of G0.253+0.016. \emph{Panel (a)} presents a cloud
  segmentation in position--position--velocity space. \emph{Panels (b,
    c)} illustrate the complex velocity structure and chemistry (note
  discrepancy between SiO and $\rm{}N_2H^+$). \emph{Panel (d)}
  presents the same information, collapsed into two velocity
  ranges. The dashed line is the lower SCUBA contour from
  Fig.~\ref{fig:sma}.\label{fig:carma}}
\end{figure*}

\subsection{SMA $\rm{}N_2H^+$ Data: Gas
  Densities $\le{}3\times{}10^5~\rm{}cm^{-3}$\label{sec:n2h+-densities}}
Figure~\ref{fig:sma}~(middle) summarizes the SMA observations of the
$\rm{}N_2H^+$ (3--2) line, observed with a beam size of
$2\farcs{}7\times{}1\farcs{}9$ (PA $46\degr{}$). It presents a
signal--to--noise ratio (SNR) map: at a given location, we divide the
signal in the brightest velocity channel by the standard deviation
obtained from channels known to be free of emission. Manual inspection
reveals emission at velocities from
$-10~{\rm{}to}~+60~\rm{}km\,s^{-1}$, indicating channels free of
emission at velocities $70$--$100~\rm{}km\,s^{-1}$. Peak positions
with an SNR $\ge{}10$ are considered as potentially detected (i.e., 56
positions); detections are deemed reliable for FWHM diameters larger
than two beams, permitting lower threshold peak SNRs $\ge{}8$ (22
positions). Figure~\ref{fig:sma} (middle) illustrates these
positions. These spectra are characterized using one--component
Gaussian fits. This approach ignores $\rm{}N_2H^+$ hyperfine blending,
which however is considered in the modeling below. Example spectra are
shown in Fig.~\ref{fig:n2h-analysis}(a). Manual inspection always
reveals one single significant velocity component per position.

We model the $\rm{}N_2H^+$ observations from
Fig.~\ref{fig:n2h-analysis}(b) using the MOLLIE non--LTE hyperfine
radiative transfer code in the hyperfine statistical equilibrium
approximation \citep{keto2010:mollie}. We adopt a relative
$\rm{}N_2H^+$ abundance of $1.5\times{}10^{-10}$ per $\rm{}H_2$ (e.g.,
\citealt{tafalla2006:internal-structure}). Spherical $\rm{}H_2$
density profiles $n=n_{0.1\rm{}pc}\cdot{}(r/0.1~{\rm{}pc})^{-2}$ are
assumed, with the density
vanishing for radii $\ge{}0.5~\rm{}pc$. The other free
parameters---i.e.\ the non--thermal gas velocity dispersion,
$\sigma_{v,\rm{}in}$, expressed by intrinsic line widths
$\Delta{}v_{\rm{}in}=(8\,\ln[2])^{1/2}\,\sigma_{v,\rm{}in}$; and the
kinetic temperature, $T_{\rm{}kin}$---are assumed to be constant
within the model sphere. We adopt $T_{\rm{}kin}=20~\rm{}K$, based on
the L12 dust temperature, resulting in optically thick (3--2)
lines. The (1--0) emission is not modelled here; it probes a larger
spatial scale not focus of the present letter. For given density,
higher abundances or temperatures imply higher intensities.

As shown in Fig.~\ref{fig:n2h-analysis}(b), the brightest
$\rm{}N_2H^+$ (3--2) peaks can be modelled using densities
$n_{0.1\rm{}pc}=(2\pm{}1)\times{}10^5~\rm{}cm^{-3}$. Integration of
the implied density profiles thus yields masses
$(260\pm{}125)~M_{\sun}$ within apertures of $0.1~\rm{}pc$ projected
radius for the most massive structures.

Surprisingly, the continuum--detected $\rm{}H_2O$--maser position is
not detected in $\rm{}N_2H^+$. The $\rm{}H_2O$--maser position is
probably less abundant in $\rm{}N_2H^+$, and thus not detectable, as
seen in some high--mass SF (HMSF) regions
\citep{fontani2006:pre-stellar-n2h+,
  zinchenko2009:hmsf-multiline}. Furthermore, all $\rm{}N_2H^+$ cores
show no significant dust emission. These cores are probably starless
and have $\rm{}N_2H^+$ abundances $\ge{}10^{-9}$, as seen in IRDCs
that resemble G0.253+0.016 in being relatively dense and starless
(\citealt{ragan2006:IRDC-lines}, \citealt{sakai2008:irdc-molecules},
\citealt{vasyunina2011:irdc-chemistry}): for example, using MOLLIE to
model cores with a higher $\rm{}N_2H^+$ abundance of $\ge{}10^{-9}$ and
$\Delta{}v_{\rm{}in}=1~\rm{}km\,s^{-1}$, the predicted $\rm{}N_2H^+$ (3--2)
line intensity is $\ge{}4.6~\rm{}K$, which is above the detection
limit (Fig.~\ref{fig:n2h-analysis}[b]), even when the density is
$n_{0.1\rm{}pc}=10^4~\rm{}cm^{-3}$, which is an order of magnitude
below that derived from the upper limit of the dust continuum flux
($<26\,M_{\sun}$ within $0.046~\rm{}pc$ radius;
Sec.~\ref{sec:dust-densities}).

\subsection{CARMA Line Emission Maps:\\ Many Fragments
  with large Velocity Differences}
Figure~\ref{fig:carma}(d) shows maps for $\rm{}N_2H^+$ (1--0) and SiO
(2--1) observed with CARMA. The beam size is
$7\farcs{}1\times{}3\farcs{}5$ (PA $6\degr{}$). We identify cloud
fragments as continuous $\rm{}N_2H^+$ (1--0) emission structures in
position--position--velocity space exceeding an intensity threshold of
$0.35~{\rm{}Jy\,beam^{-1}}=1.97~{\rm{}K}$ (noise is
$0.06~{\rm{}to}~0.12~\rm{}Jy\,beam^{-1}$). These fragments, numbered
1--7, are shown in Fig.~\ref{fig:carma}(a). The threshold was chosen
to yield a simple yet representative decomposition of the cloud
structure. Segmentation was done using 3D~Slicer\footnote{3D~Slicer is
  available from \url{http://www.slicer.org}. See
  \url{http://am.iic.harvard.edu} on astronomical research with
  3D~Slicer.} and CLUMPFIND \citep{williams1994:clumpfind}, followed
by manual removal of artifacts at map boundaries.

Fragment properties are listed in Table~\ref{tab:features}: from
spectra integrated over each fragment, we calculate
$\langle{}v\rangle{}$ and $\sigma_v$ as the intensity--weighted
velocity mean and standard deviation calculated directly from the
velocities and intensities per channel, $v_i$ and $T(v_i)$. Using
intensity--weighted mean line--of--sight velocities calculated for
every pixel, we also list the standard deviation among line--of--sight
velocities within a given fragment, $\sigma_v^{\rm{}los}$. Velocity
gradients, characterized by $\sigma_v^{\rm{}los}$, dominate the
velocity dispersion, since $\sigma_v\approx{}\sigma_v^{\rm{}los}$. The
effective radius, $R=(A/\pi)^{1/2}$, is calculated from the
CLUMPFIND--derived fragment area within the
$0.35~{\rm{}Jy\,beam^{-1}}$ intensity surface, $A$.

Figure \ref{fig:carma}(b) illustrates that the SMA--detected
$\rm{}N_2H^+$ (3--2) cores are associated with the CARMA--detected
$\rm{}N_2H^+$ fragments, as expected for cores embedded in extended
envelopes. Figure \ref{fig:carma}(c) demonstrates that the
$\rm{}N_2H^+$ fragments 4--7 are also detected in SiO.

\section{Analysis}
\label{sec:analysis}

\subsection{Star Formation Law\label{sec:sf-law}}
The most striking feature of G0.253+0.016, noted by all previous
papers on the cloud, is its low star formation rate. Here, we present
the first quantitative comparison to recently proposed ``star
formation laws''.

\citet{lada2010:sf-efficiency} suggest that molecular clouds typically
contain one embedded YSO per $\sim{}5~M_{\sun}$ of gas at $\rm{}H_2$
column densities $\ge{}7\times{}10^{21}~\rm{}cm^{-2}$. Since
G0.253+0.016 contains $2\times{}10^5\,M_{\sun}$ at column densities
$\ge{}4.5\times{}10^{22}~\rm{}cm^{-2}$ (L12, plus correction in
Sec.~\ref{sec:dust-densities}), the cloud should contain
$\sim{}4\times{}10^4$ YSOs.

\citet{lada2010:sf-efficiency} consider, of course, YSOs bright enough
to be detected. We assume that \citeauthor{lada2010:sf-efficiency}
cannot sense YSOs of mass $<0.08\,M_{\sun}$, and detect only 50\% of
stars with mass $0.08~{\rm{}to}~0.5\,M_{\sun}$. For a typical stellar
initial mass function (IMF), such as the $\alpha_3=2.7$ case of
\citet{kroupa2002:imf}, the total number of stars down to
$0.01\,M_{\odot}$ is equal to the \citeauthor{lada2010:sf-efficiency}
YSO count times a factor 2.63.

Considering this IMF, a cluster of $\sim{}4\times{}10^4$ YSOs similar
to the sources considered by \citeauthor{lada2010:sf-efficiency} would
contain stars of mass $\gtrsim{}100\,M_{\sun}$. This contradicts radio
continuum surveys for H\textsc{ii} regions, ruling out stars with mass
$\gtrsim{}16\,M_{\sun}$ in G0.253+0.016
\citep{lis1994:m0.25}. Assuming a maximum stellar mass
$\sim{}16\,M_{\sun}$, the $\alpha_3=2.7$ \citet{kroupa2002:imf} IMF,
and the factor of 2.63 to account for YSOs too faint to be detected
even in nearby clouds, the cloud should contain $\sim{}900$ YSOs of
the sort considered by \citet{lada2010:sf-efficiency}---i.e., by a
factor $\sim{}45$ lesser than the $\sim{}4\times{}10^4$ YSOs predicted
by the \citet{lada2010:sf-efficiency} law. See
\citet{lis2001:ir-spectra} for a similar IMF analysis. The
\citet{lada2010:sf-efficiency} law does thus not provide a universal
description of the SF process, contrary to assumptions by
\citet{lada2012:sf-laws} to explain the extragalactic
\citet{gao2004:hcn} infrared--HCN luminosity correlation.

\begin{table}
\caption{Fragment Properties\label{tab:features}}
\begin{center}
\begin{tabular}{llllllllllll}
\hline \hline
Fragment & $\langle{}v\rangle{}$ & $\sigma_v$ & $\sigma_v^{\rm{}los}$
& $R$ & $\alpha$ assuming\\
 & $\rm{}km\,s^{-1}$ & $\rm{}km\,s^{-1}$ & $\rm{}km\,s^{-1}$ &
   pc & $3\times{}10^{23}~\rm{}cm^{-2}$ \\ \hline
1 & $-0.1$ & 6.1 & 4.9 & 0.47 & 4.5\\
2 & 6.4 & 5.4 & 4.6 & 0.70 & 2.3\\
3 & 14.8 & 4.3 & 2.0 & 0.68 & 1.5\\
4 & 32.2 & 5.9 & 5.0 & 0.63 & 3.1\\
5 & 31.5 & 5.2 & 4.8 & 1.06 & 1.4\\
6 & 44.2 & 8.4 & 8.1 & 1.16 & 3.4\\
7 & 42.7 & 13.9 & 13.5 & 1.62 & 6.8\\
\hline
\end{tabular}
\end{center}
\end{table}

\subsection{Kinematics \& Gravitational Binding\label{sec:self-gravity}}
Stability against gravitational collapse can, e.g., be evaluated using
the virial parameter, $\alpha=5R\sigma_v^2/(GM)$ or
\begin{equation}
\alpha=1.2\,\left(\frac{\sigma_v}{\rm{}km\,s^{-1}}\right)^2
\left(\frac{R}{\rm{}pc}\right)
\left(\frac{M}{10^3\,\rm{}M_{\sun}}\right)^{-1} \, ,
\label{eq:stability}
\end{equation}
where $\sigma_v$ is the one--dimensional velocity dispersion and $G$
is the constant of gravity. Slightly depending on the equation of
state, collapse requires $\alpha{}\lesssim{}2$
\citep{bertoldi1992:pr_conf_cores, ebert1955:be-spheres,
  bonnor1956:be-spheres}.

L12 derive $M=2.0\times{}10^5\,M_{\sun}$ within $R=2.8~\rm{}pc$ (after
correction in Sec.~\ref{sec:dust-densities}), and a line width
$\Delta{}v=(8\,\ln[2])^{1/2}\,\sigma_v<16~\rm{}km\,s^{-1}$. This
yields $\alpha<0.8$: the cloud should collapse. But L12 exclude a
component at $\approx{}10~\rm{}km\,s^{-1}$ radial velocity, which our maps
show to be part of the cloud morphology (i.e., fragments 1--3;
Fig.~\ref{fig:carma}). Inclusion of the $\approx{}10~\rm{}km\,s^{-1}$
component yields $\Delta{}v=(35\pm{}5)~\rm{}km\,s^{-1}$ (Fig.~4 of
L12), resulting in $\alpha{}=3.8^{+1.2}_{-1.0}$. G0.253+0.016 thus
seems to be unbound. Fast motions $\Delta{}v>16~\rm{}km\,s^{-1}$ are
also suggested by widespread SiO shocks (Sec.~\ref{sec:nature-future}).

Still, many interferometer--detected structures seem to be
bound. Using Eq.~(\ref{eq:stability}), Table~\ref{tab:features}
reports $\alpha$ for CARMA--detected cloud fragments, assuming column
densities $\sim{}3\times{}10^{23}~\rm{}cm^{-2}$
(Sec.~\ref{sec:mass-size}). For the SMA--detected $\rm{}N_2H^+$ cores,
we adopt $R=0.1~\rm{}pc$, and the density structure from
Sec.~\ref{sec:n2h+-densities}. To include thermal pressure, we
substitute
$(\sigma_v^2+[0.188~{\rm{}km\,s^{-1}}]^2\cdot{}[T_{\rm{}kin}/10~\rm{}K])^{1/2}$
for $\sigma_v$
in Eq.~(\ref{eq:stability}). For $T_{\rm{}kin}\le{}80~\rm{}K$,
$\Delta{}v\le{}3.0~\rm{}km\,s^{-1}$, and
$n_{0.1\rm{}pc}=10^5~\rm{}cm^{-3}$ (Fig.~\ref{fig:n2h-analysis}),
$\alpha\le{}1.8$ is obtained.

Many $\rm{}N_2H^+$ (3--2) spectra reveal lines consistent with an
intrinsic line width $\lesssim{}0.5~\rm{}km\,s^{-1}$
(Fig.~\ref{fig:n2h-analysis}[b]). Compared with the
$\gtrsim{}10~\rm{}km\,s^{-1}$ lines typically found in single--dish
spectra of the GC region (e.g., \citealt{lis1998:m025}), these are
probably the most narrow lines so far detected in the GC region.

\subsection{Density Structure\label{sec:mass-size}}
\begin{figure}
\includegraphics[width=\linewidth]{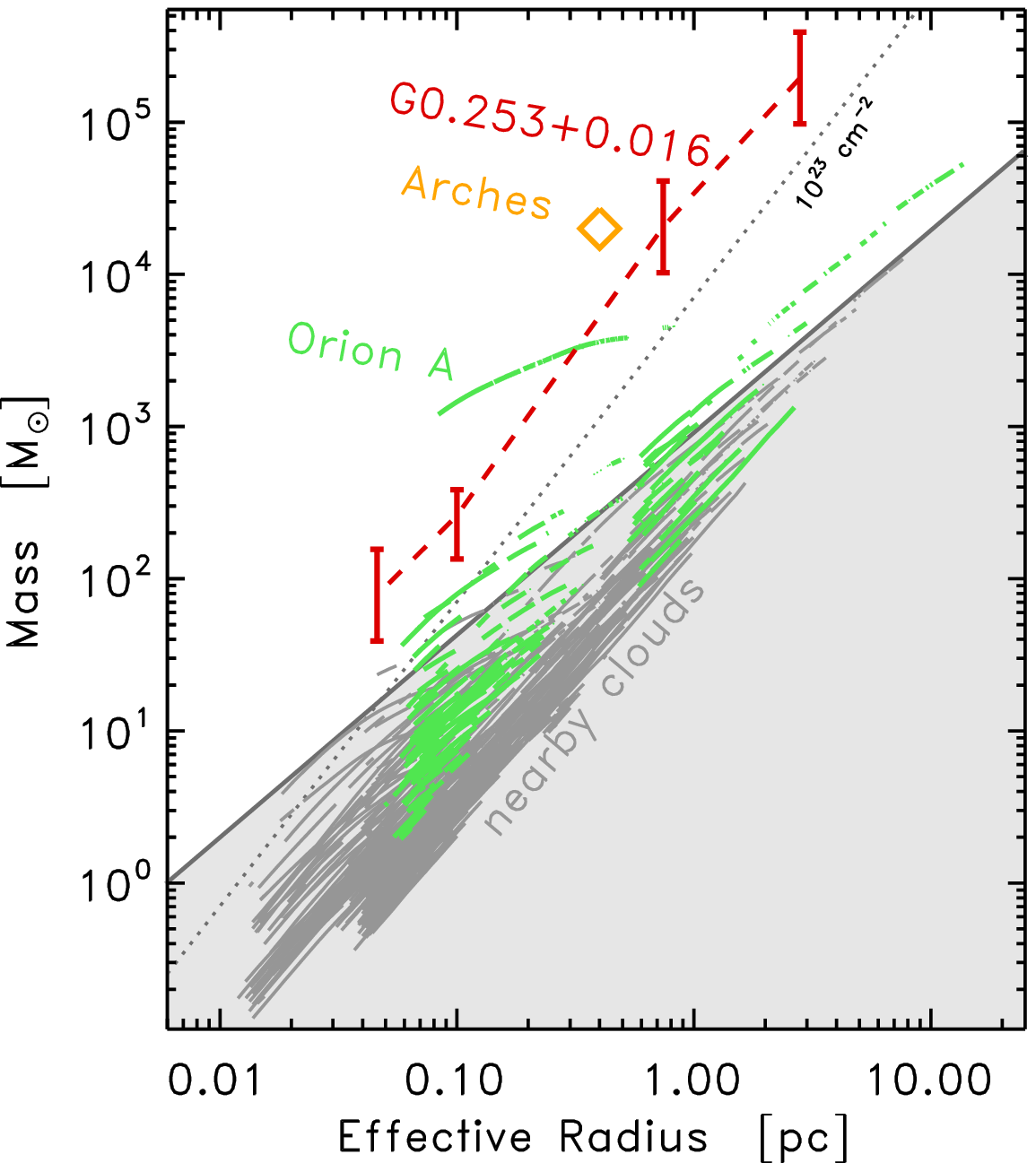}
\caption{The density structure of G0.253+0.016 (\emph{red}). Reference
  data are obtained using a hierarchical structure decomposition
  (``dendrograms'': \citealt{rosolowsky2008:dendrograms}), based on
  published structure analysis (\citealt{kauffmann2010:mass-size-i,
    kauffmann2010:mass-size-ii}; \emph{grey lines}) and previously
  unexplored Orion~A data (\citealt{kainulainen2011:confinement}, and
  see Sec.~\ref{sec:mass-size}; \emph{green lines}). For reference,
  the \emph{dotted line} highlights an $\rm{}H_2$ column density of
  $10^{23}~\rm{}cm^{-2}$. \emph{Gray line and shading} indicate the
  \citet{kauffmann2010:irdcs} limit
  (Eq.~\ref{eq:hmsf-limit}).\label{fig:mass-size}}
\end{figure}

Figure~\ref{fig:mass-size} summarizes the density structure of
G0.253+0.016. From L12, we take a mass of $2\times{}10^5\,M_{\sun}$
within $2.8~\rm{}pc$ radius, and include their peak column density of
$5.3\times{}10^{23}~\rm{}cm^{-2}$ per $36\arcsec$ beam at
$0.7~\rm{}pc$ radius ($1.7\times{}10^4\,M_{\sun}$; data scaled as
explained in Sec.~\ref{sec:dust-densities}). Interferometer--based
assessments ($78\,M_{\sun}$ and $[260\pm{}125]\,M_{\sun}$ at 0.046 and
$0.1~\rm{}pc$ radius, respectively) are from
Secs.~\ref{sec:dust-densities} and \ref{sec:n2h+-densities}. For
dust--based measurements, we adopt an opacity--induced uncertainty by
a factor 2 \citep{kauffmann2008:mambo-spitzer}. Reference data on
non--HMSF clouds are from
\citet{kauffmann2010:mass-size-ii}. Unpublished Bolocam
maps\footnote{We are indebted to D.~Li for providing the data, and
  A.~Ginsburg for reducing it.} (adopting $15~\rm{}K$ dust
temperature) and extinction data from
\citet{kainulainen2011:confinement} are used to characterize Orion~A
using methods from \citeauthor{kauffmann2010:mass-size-i}
(\citeyear{kauffmann2010:mass-size-i}, building on
\citealt{rosolowsky2008:dendrograms}). \citet{espinoza2009:arches}
characterize the Arches cluster. An approximate mass--size limit for
HMSF is taken from \citet{kauffmann2010:irdcs},
\begin{equation}
m_{\rm{}lim}(r)=870\,M_{\sun}\,(r/{\rm{}pc})^{1.33}\,{}.
\label{eq:hmsf-limit}
\end{equation}
At $r=2.8~\rm{}pc$, G0.253+0.016 exceeds the mass of equal--sized
structures in Orion~A by a factor $\sim{}25$, and the
\citeauthor{kauffmann2010:irdcs} criterion by a factor 60. The mean
$\rm{}H_2$ volume and column densities are
$M/(4/3\,\pi\,R^3)\to{}3.2\times{}10^{4}~\rm{}cm^{-3}$ and
$3.6\times{}10^{23}~\rm{}cm^{-2}$, respectively. But at smaller
spatial scales, G0.253+0.016 falls short of the masses of the Arches
cluster and the most massive structures in Orion~A by factors $\sim{}4$. At
$0.046~\rm{}pc$ radius, the \citeauthor{kauffmann2010:irdcs} criterion
is exceeded by a modest factor $\lesssim{}5$.

The interferometer--derived masses are probably underestimated. Note,
e.g., that the peak column densities from SMA and Herschel data are
similar, i.e.\ $5.2\times{}10^{23}~\rm{}cm^{-2}$ vs.\
$5.3\times{}10^{23}~\rm{}cm^{-2}$. This may result from two
factors. First, interferometer--induced spatial filtering may reduce
observed intensities. Second, the dust opacity law might be different
than assumed. None of this affects our conclusion that little dense
gas exist in G0.253+0.016. For example, if masses were higher by a
factor 5, this would imply virial parameters $\alpha\le{}0.1$ for all
SMA--detected $\rm{}N_2H^+$ cores with line widths
$\Delta{}v\sim{}0.5~\rm{}km\,s^{-1}$. Such low values for $\alpha$ are
very unusual (Pillai et al., in prep.), and thus unlikely. This
comparison suggests mass errors smaller than a factor 5.

\subsection{Decay of Gas Motions \& Accretion onto Cores}
\label{sec:decay-motions}
HMSF in G0.253+0.016 is still possible if structures in the cloud grow
more dense over time. Growth is controlled by the flow crossing time
$\ell{}/\sigma_v$ for a spatial scale $\ell$,
\begin{eqnarray}
t_{\rm{}cross}&=&1~{\rm{}Myr}\,\left(\frac{\ell}{\rm{}pc}\right)\,\left(\frac{\sigma_v}{\rm{}km\,s^{-1}}\right)^{-1}\\
&=&2.4~{\rm{}Myr}\,\left(\frac{\ell}{\rm{}pc}\right)\,\left(\frac{\Delta{}v}{\rm{}km\,s^{-1}}\right)^{-1}\,{}.
\label{eq:decay}
\end{eqnarray}
For the entire cloud, using $\ell=2R$ and
$\Delta{}v\approx{}35~\rm{}km\,s^{-1}$,
$t_{\rm{}cross}\approx{}0.4~\rm{}Myr$. Undriven turbulence is expected
to decay as ${\rm{}e}^{-t/t_{\rm{}cross}}$
\citep{maclow2004:review}. Thus, global collapse would take several
$0.4~\rm{}Myr$.

If observed velocity dispersions were reflecting pure inward motions
of speed $\sigma_v/2$, structures of constant radius $R$ could ingest
material from radii $r=R~{\rm{}to}~2R$ within the time
$(2R-R)/(\sigma_v/2)=2R/\sigma_v\equiv{}2\cdot{}t_{\rm{}cross}$. For
the $\rm{}N_2H^+$ fragments listed in Table~\ref{tab:features},
$2R/\sigma_v=0.14~{\rm{}to}~0.40~\rm{}Myr$. Adopting $R=0.1~\rm{}pc$
and $\Delta{}v=0.5~{\rm{}to}~6.0~\rm{}km\,s^{-1}$
(Fig.~\ref{fig:n2h-analysis}),
$2R/\sigma_v=0.08~{\rm{}to}~0.9~\rm{}Myr$ holds for the SMA--detected
structures.

These timescales control the structure evolution. Several
$10^5~\rm{}yr$ must pass before cores as dense as those in
Orion~A can form.

\subsection{Nature and Future of G0.253+0.016\label{sec:nature-future}}
The low SF rate for this compact and massive cloud indicates that
G0.253+0.016 is in an extreme physical state (Sec.~\ref{sec:sf-law}).
\citet{lis1998:m025} and \citet{lis2001:ir-spectra} take the existence
of widespread SiO emission as evidence for an ongoing cloud--cloud
collision. This molecule is believed to trace shocks unambiguously:
Silicon is usually locked up in dust grains, and requires grain--grain
collisions at velocities $\gtrsim{}20~\rm{}km\,s^{-1}$ to be released
\citep{guillet2009:shocks-sio}. Further gas phase reactions yield SiO
in $\lesssim{}10^3~\rm{}yr$. Figure~\ref{fig:carma} shows for the
first time that the SiO distribution is likely too smooth and extended
to result from outflows associated with a population of embedded
stars. Processes on larger spatial scales, such as cloud--cloud
collisions, are a more probable origin. It thus seems plausible that
G0.253+0.016 is a very young cloud that will soon dissipate internal
motions and efficiently form stars in a few $10^5~\rm{}yr$
(Sec.~\ref{sec:decay-motions}).

However, the cloud may not be gravitationally bound and simply
disperse (Sec.~\ref{sec:self-gravity}). Furthermore, G0.253+0.016 is
subject to the disruptive GC environment: as already mentioned by L12,
following the GC orbit proposed by \citet{molinari2011:cmz-ring},
G0.253+0.016 will arrive at the present location of Sgr~B2 in
$\sim{}8.5\times{}10^5~\rm{}yr$. The latter cloud essentially
represents a standing shock, where gas clouds on different GC orbit
families collide (e.g., \citealt{bally2010:bolocam-gc}). Given the
disturbed nature of the Sgr~B2 region, it is not clear whether
G0.253+0.016 will then be disrupted or be pushed into collapse.

\section{Conclusion}
\label{sec:conclusion}
G0.253+0.016 deviates from current ``star formation laws'' (e.g.,
\citealt{lada2010:sf-efficiency}) by a factor $\sim{}45$
(Sec.~\ref{sec:sf-law}). The scarcity of significant dust and
$\rm{}N_2H^+$ cores in our SMA interferometer maps
(Secs.~\ref{sec:dust-densities}--\ref{sec:n2h+-densities}) reveals
that G0.253+0.016 is presently far away from forming high--mass stars
and clusters (Sec.~\ref{sec:mass-size}): considerable evolution for
several $10^5~\rm{}yr$ is needed before such star formation might
occur (Sec.~\ref{sec:decay-motions}). The cloud might thus be very
young and currently forming in a cloud--cloud collision indicated by
SiO shocks (Sec.~\ref{sec:nature-future}). Given the disruptive
dynamics of the Galactic Center region (Sec.~\ref{sec:nature-future}),
and the potentially unbound nature of the cloud
(Sec.~\ref{sec:self-gravity}), it is unclear whether evolution towards
significant star formation will ever happen.

\acknowledgements{We thank S.~Longmore for giving us access to
  \citet{longmore2011:m025} in advance of publication, D.~Lis and
  K.~Menten for enlightening discussions, and an anonymous referee for
  making the paper more readable. JK is grateful to D.~Li and
  P.~Goldsmith, his hosts at JPL, for making this work possible. Part
  of the research was carried out at the Jet Propulsion Laboratory,
  California Institute of Technology, under a contract with the
  National Aeronautics and Space Administration. TP acknowledges
  support from CARMA, supported by the National Science Foundation
  through grant AST~05--40399.}

Facilities: \facility{SMA}, \facility{CARMA}, \facility{Spitzer}


\end{document}